\documentclass[a4paper,aps,prl,twocolumn]{revtex4}

\usepackage{graphicx} 
\usepackage[usenames,dvipsnames]{xcolor}
\usepackage{tikz}
\usepackage{amsmath}
\usepackage{amssymb}
\usepackage{natbib}
\usepackage{color}
\usepackage{psfrag}
\usepackage{wasysym}
\usepackage{float}
\usepackage{epstopdf}
\usepackage[normalem]{ulem}
\usepackage{pgfplots}
\usetikzlibrary{pgfplots.groupplots}
\usepackage{textcomp}
\usepackage{hyperref}
\usepackage{soul}
\usepackage{amsmath}

\pgfplotsset{compat=newest}
\pgfplotsset{yticklabel style={text width=1.3em,align=right}, xticklabel style={text width=1.5em,align=center}}
\pgfplotsset{every axis legend/.append style={anchor=south east}}

\begin{document}
\title{Emergence of Structure in Columns of Grains and Elastic Loops}
\author{Arman Guerra and Douglas P. Holmes }
\affiliation{\footnotesize Mechanical Engineering, Boston University, Boston, MA, 02215, USA}

\date{\today}

\begin{abstract}
It is possible to build free-standing, load-bearing structures using only rocks and loops of elastic material. We investigate how these structures emerge, and find that the necessary maximum loop spacing (the critical spacing) is a function of the frictional properties of the grains and the elasticity of the confining material. We derive a model to understand both of these relationships, which depends on a simplification of the behavior of the grains at the edge of a structure. We find that higher friction leads to larger stable grain-grain and grain-loop contact angles resulting in a simple function for the frictional critical spacing, which depends linearly on friction to first order. On the other hand, a higher bending rigidity enables the loops to better contain the hydrostatic pressure of the grains, which we understand using a hydroelastic scale. These findings will illuminate the stabilization of dirt by plant roots, and potentially enable the construction of simple adhesion-less structures using only granular material and fiber.
\end{abstract}


\maketitle

Ensembles of dry, adhesion-less grains can flow like a fluid~\cite{lun1984kinetic, silbert2001granular, kamrin2012nonlocal}, or jam into a solid-like state~\cite{majmudar2007jamming, song2008phase}. There are two notable mechanisms by which grains jam. The first is the densification induced by an external force~\cite{liu1998jamming, van2009jamming}, {\em e.g.} coffee beans will jam when vacuum packed in an air-tight bag. Jamming occurs when the number of inter-grain contacts reaches a critical value -- the aggregate becomes isostatic, since each grain is held in place by its neighbors or the container~\cite{van2009jamming}. As anyone who has run their hand through the sand on a beach knows, however, these states are fragile, and a small change in the direction of the external forces can often induce flow. The second mechanism is the entanglement or interlocking of particles. This occurs either when the particles have a high aspect ratio and are flexible, such as when birds build a nest made from sticks~\cite{hansell2000bird, weiner2020mechanics}, fibers are spun into felt or cotton balls~\cite{kabla2007nonlinear, picu2011mechanics, gravish2012entangled}, or when particles geometrically interlock, such as when ants assemble themselves into bridges and rafts~\cite{anderson2002self}, or staples are mixed together~\cite{franklin2012geometric, murphy2016freestanding}. These jammed states can be very stable in the absence of any additional, external confinement. 

Both of these mechanisms appear when plant roots penetrate into soil, {\em i.e.} an \textit{elastogranular} interaction~\cite{schunter2018elastogranular} between a slender elastic structure and a granular material. The elastica will encircle and confine grains, and at the same time entangle with themselves, securing themselves and the grains that they contact in place~\cite{reubens2007role, ghestem2014influence}. Entangled plant roots aid in preventing landslides and stabilizing the banks of rivers~\cite{gray1981forest, nilaweera1999role}. Recently this method of stabilizing granular matter has been used to build structures -- researchers have combined different kinds of fibers and grains and shaped them into load bearing walls and columns~\cite{aejmelaeus2016jammed, aejmelaeus2017granular, Cohen2020}.

 \begin{figure}
\begin{center}
\vspace{1mm}
\includegraphics[width=0.98\columnwidth]{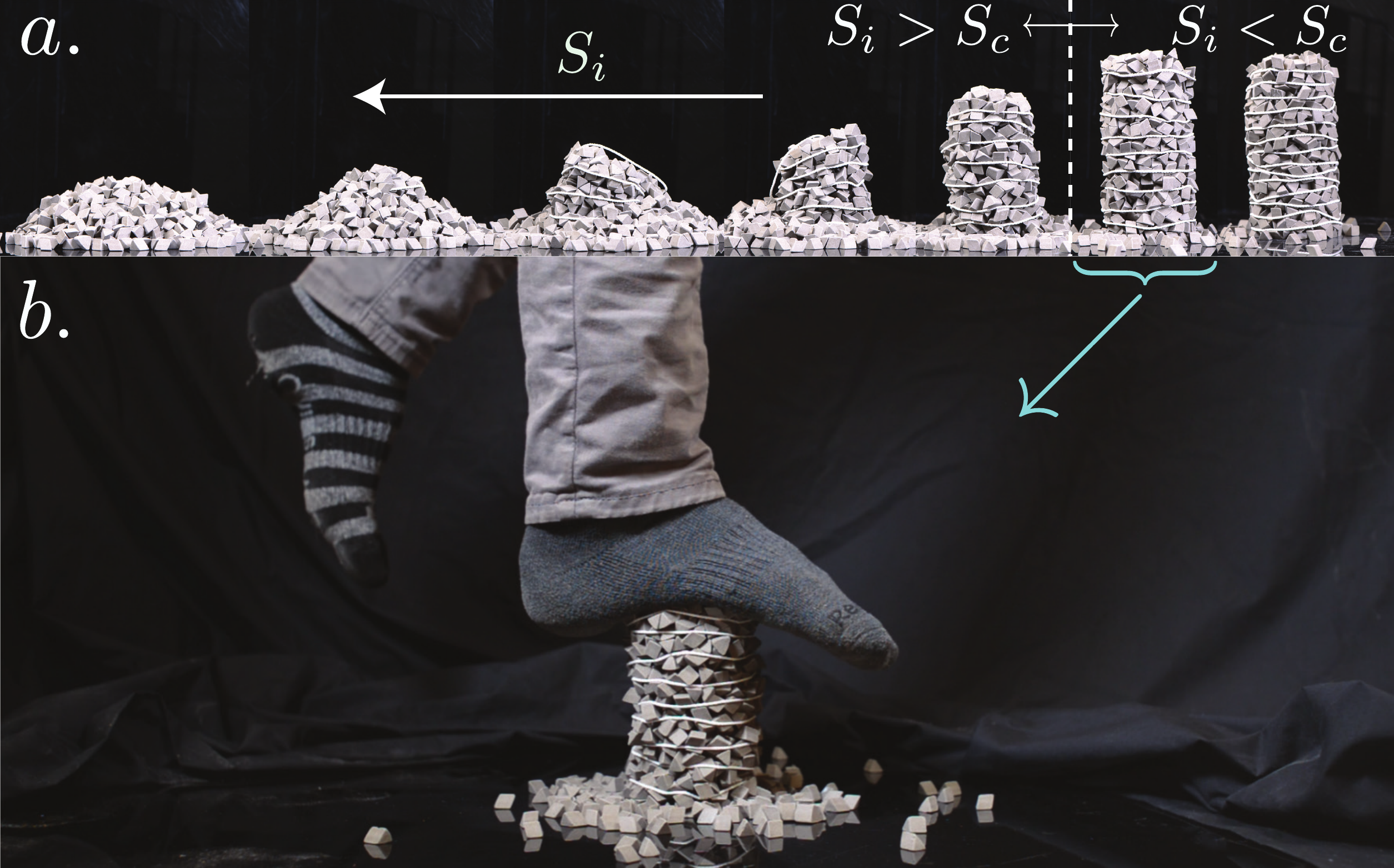}
\end{center}
\vspace{-7mm}
\caption{ (a) The emergence of structure of a column of grains with exterior loops of string. Each frame is a column prepared with a different initial spacing $S_{i}$. The frame indicated with the blue curly bracket is prepared with $S_i$ just below the critical spacing $S_{c}$. (b) The column underlined in blue in (a) loaded to over $10^5$ Pa
\label{Fig1}}\vspace{-5mm}
\end{figure}

In this Letter, we consider a simplified version of these structures, columns made from grains and loops of elastic rods that provide external confinement (Figure~\ref{Fig1}). These loops act as containers for the grains, but in contrast to many other structures made from confined grains~\cite{wu2008behavior, ramli2013stability}, the spacings between the loops can be larger than the grain diameter. Therefore, we ask the question: what are the minimum criteria to form a stable, elastogranular column?

\section{Methods}

We will parameterize the positions of the loops as the spacing between them, $S$. In the limit of very large initial loop spacing ($S_{i} \rightarrow \infty$), a column will collapse into a pile, the shape of which is determined by the properties of the grains (Figure~\ref{Fig1}a, left)~\cite{al2018review}. As $S_{i} \rightarrow 0$, the column will retain most of its initial shape. We define the critical spacing $S_{c}$ as the maximum loop spacing that will allow the columns to stand up with minimal reduction in their height (schematic in Figure~\ref{big}a) {\em i.e.} the final height $H_{f}$ is within $95\%$ of $H_{i}$ (to account for grain settling, more detail in Appendix A). When $S_{i}$ crosses below $S_{c}$, the column may still lose {\em rattler} particles, {\em i.e.} particles which do not contribute to the stability of the jammed state~\cite{baule2018edwards}, but retain its initial imposed shape. Decreasing $S_{i}$ further leads to no qualitative change in the final shape (Figure~\ref{Fig1}a, right). We note that a column prepared at $S_{i} \approx S_{c}$ can handle a uniaxial compressive stress of over $10^5$ Pa (Figure~\ref{Fig1}b).

\subsection{Experiments}

To determine $S_{c}$, we built elastogranular columns which varied in $S_{i}$, and considered a wide variety of granular matter (glass beads, plastic sous-vide balls, peanut M\&M's, and ceramic rocks --  diameters ranging between 1.0 and 1.9 cm) which we selected based on their varying frictional properties. We note that all of the granular material we considered is approximately spherical making it otherwise challenging to stabilize, {\em i.e} we do not consider any grains with large aspect ratios~\cite{weiner2020mechanics}, interlocking parts~\cite{murphy2016freestanding} {\em etc.}, which may otherwise become kinematically trapped into a structural form. First, we poured the granular particles into an 8 cm diameter hollow, hard, cylindrical slip-cast mold (Clear Cast Acrylic Tube, 3-1/2" OD x 3-1/8" ID, McMaster) until they reached a prescribed initial height $H_{i}$ of 16cm. At regular intervals in this pouring process we leveled the grains and placed an 8cm diameter loop around the exterior of the grains. For our first experiments we used string (type 18 Twisted Mason's Twine, McMaster). We then removed the slip cast mold vertically in a quasi-static manner (20mm/s) using a linear actuator (Zaber Technologies T-LSR300B), and measured the final height $H_{f}$ of the elastogranular column when it came to rest. We use the angle of repose $\alpha$, commonly defined as the angle that a quasi-statically heaped pile of grains makes with the ground~\cite{al2018review}, to account for the friction between grains, as well as any slightly non-spherical geometrical features. In the case that the grains are made from an ideal, cohesion-less Coulomb material, this angle is related to the coefficient of friction by $\alpha = \phi = \arctan{\mu}$, where $\phi$ is the angle of internal friction and $\mu$ is the coefficient of static sliding friction~\cite{nedderman2005statics}. This assumption is not perfect -- in reality $\alpha$ is a complicated function of the rolling and sliding friction, gradation, shape, {\em etc.} of the particles~\cite{zhou2002experimental, robinson2002observations, al2018review}. There are in fact many definitions and ways to measure the shear strength and frictional properties of granular materials, which may be applicable to different particle sizes, shapes, gradations, and loading scenarios~\cite{al2018review, chakraborty2010dilatancy}. However, since our grains have a comparably high sphericity and are of uniform size and shape, and further since they are under a self-load on a flat surface, we believe that the angle of repose is a sufficient and appropriate metric to estimate the role of friction and geometry in this work.

\begin{figure}
\begin{center}
\vspace{1mm}
\includegraphics[width=0.98\columnwidth]{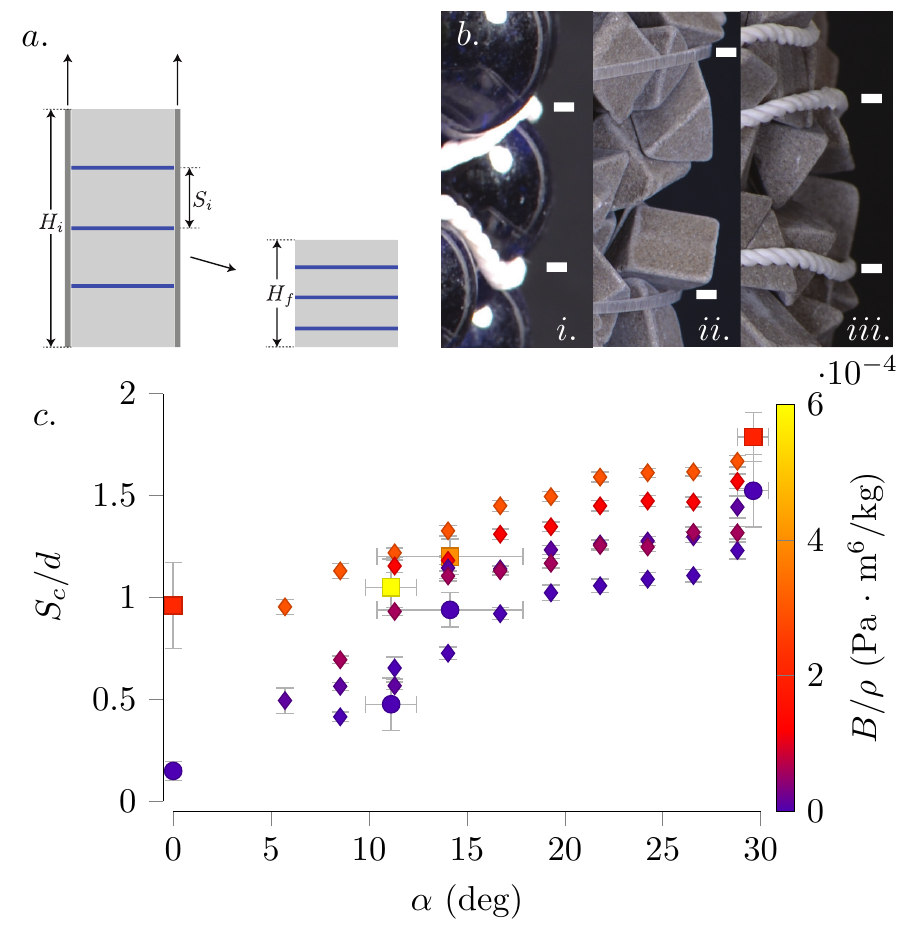}
\end{center}
\vspace{-7mm}
\caption{ (a) A schematic of our experiment. We build columns inside of an acrylic slip-cast mold to a height $H_{i}$, then quasi-statically remove the mold and measure $H_{f}$. The critical spacing $S_{c}$ is defined as the minimum spacing such that the column does not reduce in height (rigorously defined in Appendix A). (b) Edges of columns made from (\textit{i}) Marbles ($\alpha=0$) and string, (\textit{ii}) Ceramic ($\alpha\sim30$) and acrylic rings, and (\textit{iii}) Ceramic and string. White lines are added to draw the eye to the loops (c) $S_{c}$ vs $\alpha$ for experiments (strings-- circles, acrylic -- squares) and simulations (diamonds).  Data is colored by the bending rigidity of the loops ($\sim Eh^3$) divided by the density of the grains $B/\rho$ (Pa $\cdot$ m$^6$/kg). The yellow data point has $B/\rho \sim 30 \times 10^{-4}$ but we have cut off the colorbar for visualization.
\label{big}}\vspace{-5mm}
\end{figure}

Intuition would suggest that if the grains are approximately spherical, $S_{c}$ would be on the order of the grain diameter $d$, and that $S_{c}$ will increase with the grain--grain friction. However, the strings are flexible, in addition to being able to translate and rotate, and as such we find that if the friction between the grains is low, the grains tend to push the string out of the way and escape even when $S_{i} < d$ (Figure~\ref{big}b, (\textit{i})). In the limit of large $\alpha$, we find that grains will tend to collocate into stable arrangements between strings (Figure~\ref{big}b (\textit{iii})), allowing $S_{c}$ to reach as high as $1.5d$. We plot $S_{c}/d$ vs $\alpha$ for our experiments with string in  Figure~\ref{big}c (circles).

For the same range of granular material properties, we also investigated the opposite limit -- where the strings are replaced with rigid acrylic rings of the same diameter and thickness (Figure \ref{big}c, squares). We found that $S_{c}$ for these experiments followed a similar trend in $\alpha$, and was always higher than experiments using strings. We attribute the difference between the behavior of columns made with string and with acrylic to the negligible bending rigidity of the string, which is noted in Figure~\ref{big}b (\textit{ii}) -- the grains do not bend acrylic rings much, and as such the grains can be stable even when they only overlap with the rings slightly. On the other hand when the loops are flexible, the grains tend to bend them out of the way, thereby providing a means for escape. One might expect that if the grains are perfectly frictionless they may slip between the flexible strings no matter how small the spacing is, whereas no rigid grain can move between acrylic loops which are spaced less than the grain diameter. 

\subsection{Numerical Simulations}

We next investigated the influence of bending rigidity of the confining loops on the stability of the columns. Experimentally it is difficult to vary the bending rigidity of the confining loops without changing their material and geometric properties. To complement our experimental data, we repeated the experiments in the Large-scale Atomic/Molecular Massively Parallel Simulator (LAMMPS)~\cite{plimpton1995fast} using hard (Young's Modulus $E_g = 10^{9}~\mathrm{Pa}$, Poissons ratio $\nu = 0.4$, coefficient of restitution $(e)=0.35$) spheres (diameter $d$=1.15 cm). We use tangential and rolling stiffnesses $K_t = 4 E_g/2(2-\nu)(1+\nu)$, $K_r=0.1K_t$ and rolling damping coefficient $\gamma_r = 1$~\cite{jiang2005novel, luding2008cohesive, horabik2016parameters}. We varied the sliding friction coefficient $\mu_{s}$ from 0.1-0.55 and we took the rolling friction coefficient $\mu_{r} = \mu_{s}$~\cite{jiang2005novel, luding2008cohesive, horabik2016parameters}. To make confining loops we simulate many small spheres (Figure~\ref{collapsefig}e) with nearest-neighbor potentials 

\begin{equation}
U = Y(r-r_{0})^2 + B(1+\cos(\theta))
\end{equation}

Which includes a stretching term, a harmonic function of the distance between adjacent loop particles $r$ and the equilibrium distance $r_{0}$, and a bending term, which is a function of the angle between groups of three respective loop particles $\theta$. If we take the stretching modulus $Y = E_l \pi h /8$ and the bending modulus $B = E_l \pi h^3/64$ where $E_l$ is the Young's modulus of the loops and $h$ is the thickness of the confining loop, we will recover the formula for the energy of a cylindrical elastic loop (Appendix B). We fix the stretching rigidity and vary the bending rigidity of the loops in simulations (Figure~\ref{big}c, diamonds). We will note that, in addition to enabling us to systematically vary the bending rigidity of the loops, the simulations were complementary to our experiments in that they allowed us to study an experimentally inaccessible range of $\alpha$, specifically $\alpha \in (1, 10) \cup (16, 26)$, for which were unable to find grains in our target diameter range, and enabled a systematic study of the dependance of the bending of the loops on each experimental parameter, thereby helping to justify a scaling law for the system (see Figure~\ref{collapsefig}).

\section{Analysis}

These experiments and simulations so far indicate that the critical spacing at which columns stand up is a function of the angle of repose $\alpha$ of the grains and the bending rigidity $B$ of the confining loops. To understand how these factors play a role, we will consider a reduced-order model of the grain arrangement at the edge of a column (which takes into account the possible contributions of grain-grain friction, grain-loop friction, and loop bending) and find the conditions for local stability. This simplification will allow us to establish some guidance on what combination of material and geometric parameters will enable the emergence of a column that can bear its own weight.

\begin{figure}
\begin{center}
\vspace{1mm}
\includegraphics[width=0.98\columnwidth]{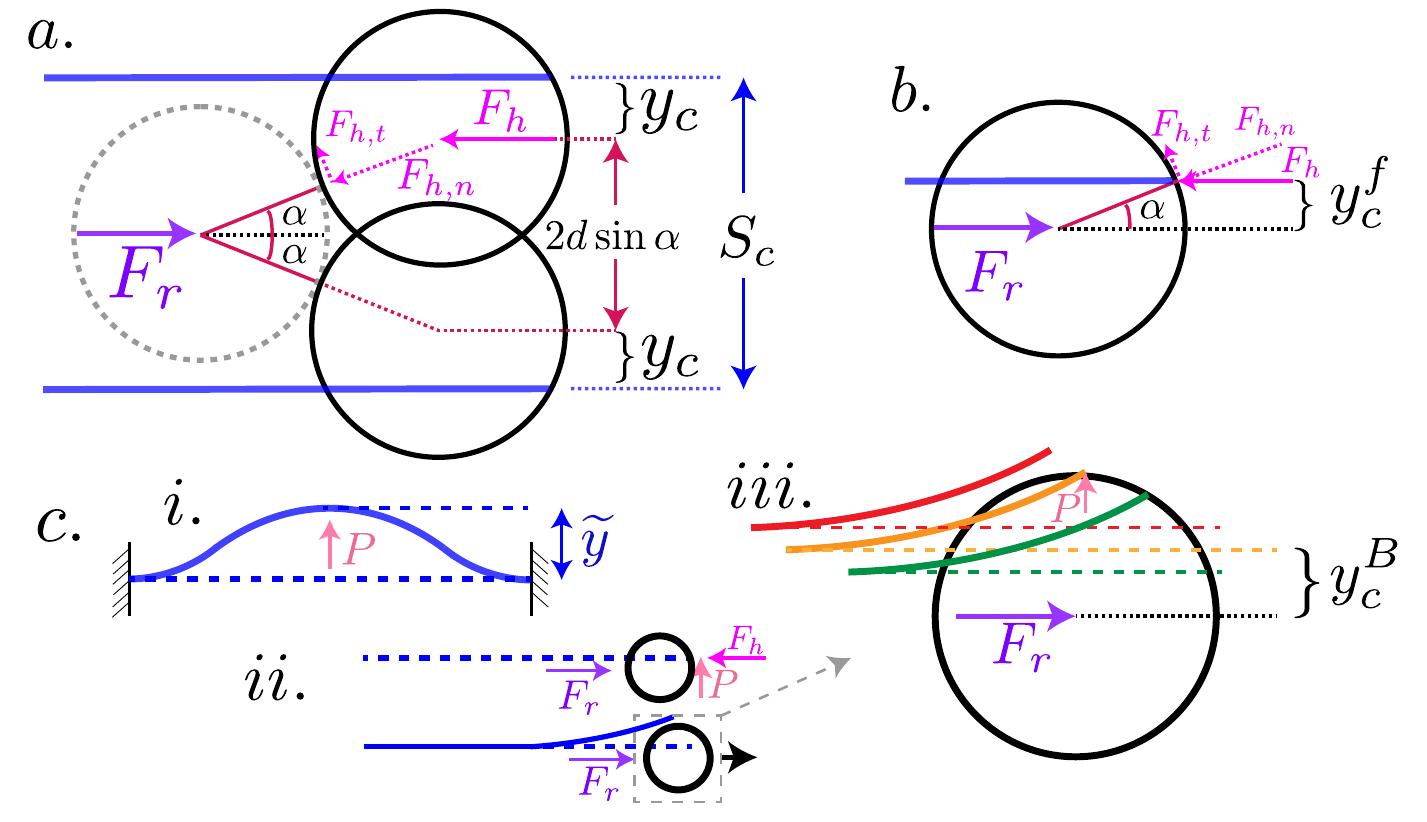}
\end{center}
\vspace{-5mm}
\caption{ (a) A schematic of our mathematical model. An interior grain is subject to an outward radial force but held in place by two exterior grains which contact the loops. According to the Coulomb constraint, for the grain-grain contact point to be stable, the angle between the centers of the interior and exterior grains can be at most $\alpha$, so the maximum stable distance between loops $S_c = 2d\sin{\alpha} + 2y_c$. (b) We denote $y_c^f$ as the maximum stable height of the loops with respect to the grains due to frictional stability, which can be found with a similar constraint as above. (c) (\textit{i})  An elastic rod loaded with a force $P$ will bend a distance $\widetilde{y}$ proportional to the ratio of the applied force and the bending modulus. (\textit{ii})  The radial force from the interior grains threatens to bend the loops out of the way. (\textit{iii}) We denote $y_c^B=d/2-\widetilde{y}$ as the maximum stable height of the loops with respect to the grains due to the bending rigidity of the loops.
\label{modelfig}}\vspace{-5mm}
\end{figure}

Consider an arrangement of three grains near the edge of a column, one interior grain contacting two exterior grains both of which contact the confining loops (Figure~\ref{modelfig}a). We will assume that the grains are spherical and monodisperse, and we note that since we have illustrated a 2 dimensional lateral cross section of a 3 dimensional scenario, the exterior grains may vertically overlap with one another. In the absence of the exterior grains, the interior grain would escape the column radially, and as such there must be an outward radial force acting on the interior grain which is balanced by the exterior grains. Consequently, there must be a force acting inwards on the exterior grains from the confining loops -- to understand the role of grain-grain friction we will first assume that this force results from a hoop stress, that is, it acts radially inwards. We illustrate the outward radial force as $F_r$ and the inward hoop force as $F_h$.

From here, we can determine whether the contact point between the interior and exterior grains will be stable or will slip based on their relative positions. Using the familiar Coulomb constraint, we can say that the contact will be stable if $\mu F_{h,n} \geq F_{h,t}$ where $F_{h,n}$ and $F_{h,t}$ are the components of the hoop force that are normal and tangential, respectively, to the tangent line of contact between the spheres, illustrated in Figure~\ref{modelfig}a. This provides an upper bound for the angle $\theta$ between the horizontal and the vector which points from the centers of the interior and exterior grains -- we can say $\alpha = \arctan{\mu} \geq \arctan{F_{h,t}/F_{h,n}} = \arctan{F_h\sin{\theta}/F_h\cos{\theta}} = \theta$. Therefore the maximum distance between the centers of the exterior grains is $2d\sin{\alpha}$.

The only further unknown is the maximum vertical distance between the loops and the centers of the exterior grains. To find this we will separate out the influence of the friction from the influence of the bending rigidity -- one could imagine that if the loops had negligible bending rigidity, but there was a high degree of friction between the grains and the loops, the loops may constrain the grains because the contact point between the loops and the grains does not slip. Alternatively, in the case of infinitely slippery grains and loops, if the loops were stiff, the grain-loop contact point may slip but the loops may not bend out of the way enough for the grains to escape. We will call $y_c^f$ the maximum critical height of the loops due to frictional stability, and $y_c^B$ the maximum critical height of the loops due to bending rigidity. We will then find two versions of the largest stable spacing, $S_c^f = 2d\sin{\alpha}+2y_c^f $ and $S_c^{B} = 2d\sin{\alpha}+2y_c^B$. The actual critical spacing will be the overall largest stable spacing, and therefore is the larger of the two -- $S_c/d = \mathrm{max}(S_c^f/d,S_c^B/d)$

 \begin{figure}
\begin{center}
\vspace{1mm}
\includegraphics[width=0.98\columnwidth]{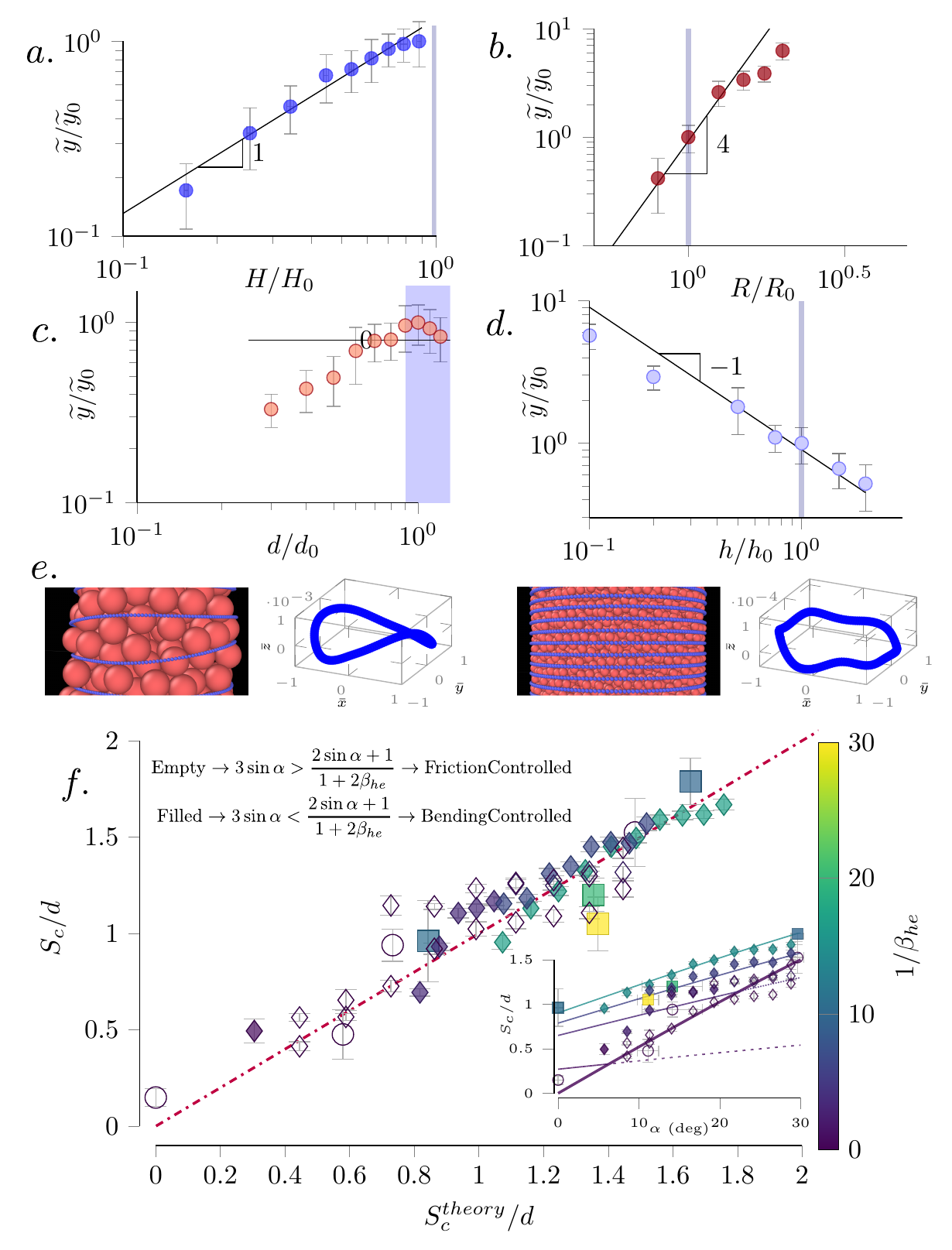}
\end{center}
\vspace{-5mm}
\caption{ (a) We fix $\alpha \sim 17$deg and measure the change in $\widetilde{y}$ as we individually vary the height of the columns, (b) radius of the columns, (c) diameter of the particles and (d) thickness of the loops in simulations. We normalize each input variable and $\widetilde{y}$ by the values in the original simulations ($H_0, R_0, d_0, h_0, \widetilde{y}_0$). The black guide lines have the slopes from our mathematical model: 1, 4, 0, and -1 respectively, and the blue lines indicate the range of the parameters in experiment. (e) Two images from our simulations,~\cite{stukowski2009visualization} and a 3 dimensional graph of a ring selected from each. The z-axis is scaled up to emphasize the curvature of the ring. The bending length $L$ of the rings is a function of both $R$ and $d$. We ignored the $d$ contribution, which is why we overestimate the dependance of $\widetilde{y}$ at high $R$ and underestimate at low $d$. (f) Comparison between $S_c^{theory}/d = \mathrm{max}(3\sin{\alpha} , \frac{2 \sin{\alpha}+1}{1+2 \beta_{he}})$ and our experimental values of $S_c$. We color by the inverse of the hydroelastic scale, $1/\beta_{he}$ and fill based on whether experiments are in a friction controlled or bending controlled region of $(\alpha,\beta_{he})$ space. The yellow data point has $1/\beta_{he} \sim 150$ but we have cut off the colorbar for visualization. Error bars are taken from Figure~\ref{big} and do not include the error in $x$. In the inset we show the data from Figure~\ref{big}c with our analytical results plotted. The thicker line represents $S_c^f$ and the thinner lines represent $S_c^B$ for different $\beta_{he}$. The dotted sections represent when $S_c^B$ is below $S_c^f$. 
\label{collapsefig}}\vspace{-5mm}
\end{figure}

We start with $y_c^f$. If we still consider the hoop force acting between the exterior grains and the loops, and we assume that the grain-grain coefficient of friction is the same as the grain-loop coefficient of friction, we find ourselves with an almost identical constraint as before -- the maximum value of the angle between the horizontal and the vector pointing between the centers of the grains and the loops is $\alpha$ and we find that $y_c^f=\frac{d}{2}\sin{\alpha}$

Now, to find $y_c^B$ we must consider the force which acts to bend the loop out of the way. We will set aside the grain-loop friction and relax the hoop-force assumption, such that now there is a vertical force between the grains and the loops. When a force $P$ acts on an Euler--Bernoulli beam with a cylindrical cross-section $h$, length $L$, bending modulus $B$, and clamped edges, it will bend with a maximum amplitude 

\begin{equation}
\widetilde{y} = \frac{P L^{3}}{192 B h}
\end{equation}

Illustrated in Figure~\ref{modelfig}c (\textit{i}). If the loop is to constrain a grain, then it must not bend out of the way so much as to clear the top of the grain (Figure~\ref{modelfig}c (\textit{ii})). We can therefore take the height of the top of the grain $d/2$ and subtract the bending of the loop $\widetilde{y}$ to find the maximum height of the loop such that it will not bend out of the way of the particle, $y_{c}^B = d/2 - \widetilde{y}$ illustrated in Figure~\ref{modelfig}c (\textit{iii}). To put $y_{c}^B$ in terms of the variables of our problem, we will assume that the force on each loop is due to the hydrostatic pressure from the grains, which will be $ \phi_{rlp} \rho g D$ at a depth $D$, density $\rho$, and random loose packing fraction $\phi_{rlp} \sim 0.55$. The area associated to each loop is $2\pi R S_{c}$ where R is the radius of a column. Therefore the average force on each loop is $P \sim \phi_{rlp} \rho g (H/2) 2\pi R S_{c}$ where $H$ is the height of a column. Depending on how the force is distributed on each loop, the bending length L will either scale with the total length of the loop $\sim R$ or on the fluctuations in the force applied to the loop $\sim d$. We find qualitatively from our simulation data that, in the range of $d/R$ that we study, most of the force on a given loop can be attributed to a small number of grains, which implies that $L$ will be limited by $R$. This gives the result

\begin{equation}
y_{c}^B = \frac{d}{2} - \frac{\phi_{rlp}\pi}{192}\left( \frac{\rho g H R S_{c}}{B h/R^3}\right)
\label{yc}
\end{equation}

Where we have expressed the second term as a ratio between the hydrostatic force on the loops and their bending rigidity. We will separate $S_c$ from this second term and wrap the rest of the term into a {\em hydroelastic} scale, $\beta_{he} = \frac{\phi_{rlp}\pi}{192}\left( \frac{\rho g H R }{B h/R^3}\right)$.

We note that in our experiments and simulations so far the only variables which have changed significantly are $S_c$, $\rho$, and $B$. To further test our scalings of $\widetilde{y}$ in the variables that we had not yet varied, we performed some additional simulations, the results of which are shown in Figure~\ref{collapsefig}. We note that when $d/R$ is small, $\widetilde{y}$ starts to scale with $d$ and the $R$ dependence reduces. This occurs because as the number of contacts with the loop increases, the force fluctuations start to limit the bending of the loops, and the bending wavelength shortens (Figure~\ref{collapsefig}e).  But, within our experimental range (marked in blue), our scalings seem appropriate.

We can now plug in our values of $y_c^f$ and $y_c^B$ to find the critical spacing. Rearranging, we find that

\begin{equation}
\begin{split}
& S_c^f/d =  3\sin{\alpha}  \\ 
& S_c^B/d = \frac{2\sin{\alpha}+ 1}{1+2\beta_{he}}
\label{Sc}
\end{split}
\end{equation}

The expected stability is therefore set by the maximum of these two values, one which is solely dependent on the friction ($\alpha$) and a second that is dependent on both friction and the bending rigidity of the loops ($\alpha$ and $\beta_{he}$ where $\beta_{he} \propto B^{-1}$, see Equation~\ref{yc}). We plot $S_c/d$ against our theoretical finding in Figure~\ref{collapsefig}f and find good agreement between our experiments and the behavior derived from our simplified model, indicating that this model provides good intuition for the interactions at play. This reduced order model describes the minimum conditions necessary for a column of loops encased by a finite number of elastic loops to retain its shape when placed in a gravitational field. However, the simplifications in the model make it difficult to comment on the stability of the resulting equilibrium shapes. Indeed, these columns may be fragile to eccentric loading, shear, or bending, and may be sensitive to defects and imperfections which are effectively averaged over by our inherent assumptions. These are important considerations left to future work, however we reiterate that, once these structures are jammed in response to uniaxial loading, they are capable of bearing significant loads (Figure~\ref{Fig1}).

\section{Adaptable Forms}

 \begin{figure}
\begin{center}
\vspace{1mm}
\includegraphics[width=0.98\columnwidth]{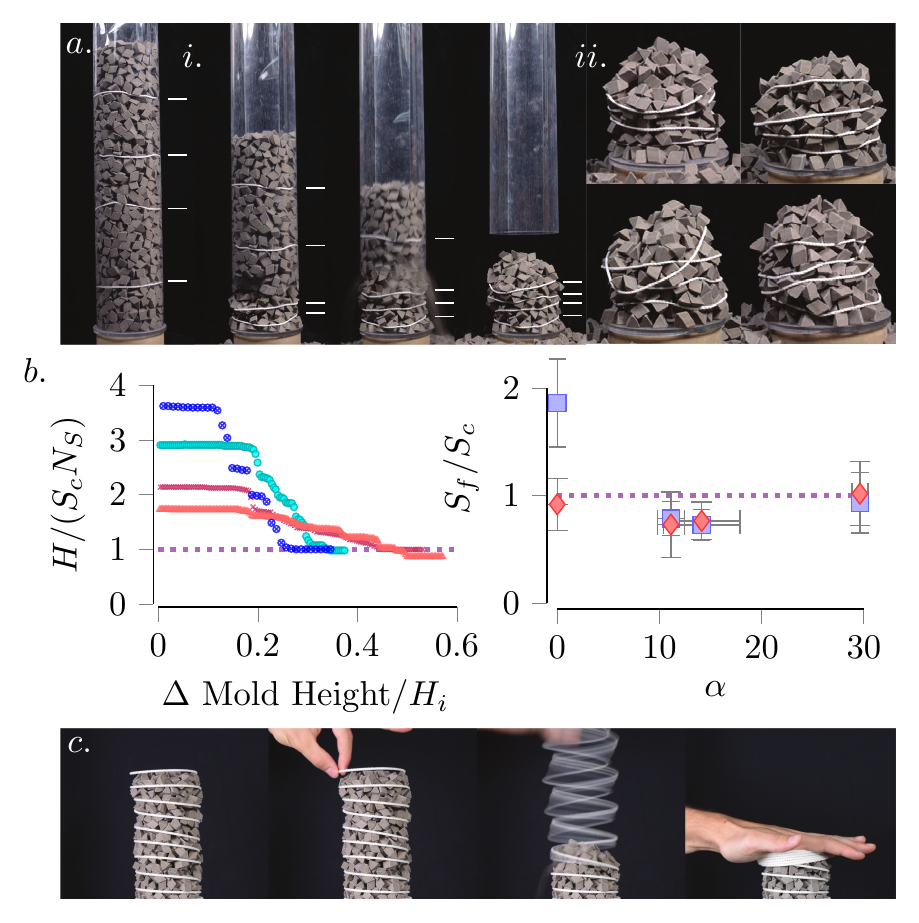}
\end{center}
\vspace{-5mm}
\caption{ (a) (\textit{i}) Snapshots of an experiment with ceramic grains where columns are built with an initial height of 33cm and $S_i \sim 3.6 S_{c}$. (\textit{ii}) The final state of the columns for different runs of the experiment shown in (\textit{i}). As grains escape between the strings, the columns shrink until they end up at a $H_{f}$ which depends on the number of spacings $N_{S}$ (which includes the spacing between the bottom (top) of the column and the bottom (top) string). (b) Left: Height of columns made from ceramic grains as the slip cast-mold is removed. We normalize the height by the product $S_c N_{S}$. If each spacing was exactly $S_{c}$ this value would be 1. Right: We find $S_{f}$ at multiple points around the circumference of the cylinder and plot the average (blue circles) and median (red diamonds) value, normalized by $S_{c}$, for different $\alpha$. (c) We can use these ideas to make structures that are stable in compression and can be dismantled by a simple user input.
\label{spacingfig}}\vspace{-5mm}
\end{figure}

Now finally we ask the question: what happens when we prepare columns with $S_{i}>S_{c}$? When constructed on flat ground, the columns become encompassed by a pile of the grains which escape, as shown in Figure \ref{Fig1}a. This pile obscures the final state of the loops. If instead, we set the columns on a pedestal that is the same diameter as the diameter of the columns, the escaping grains fall away and we can observe the dynamics of the loops. 

The evolution of the average spacing of strings $S$ for various columns with $S_{i}>S_{c}$ is shown in Figure \ref{spacingfig}b. Remarkably, as the columns settle, $S\rightarrow S_c$. In other words, a string will fall until the spacing between it and the string below it (or the ground) is, on average, \emph{equal to} $S_{c}$, irrespective of the initial spacing. The result is a shorter column with a final spacing $S_{f}=S_{c}$. A striking example of this phenomenon is shown in Figure \ref{spacingfig}a. This behavior emphasizes the robustness of a $S_{c}$ which is impervious to dynamical effects, and indicates the possibility that this technology could be used to construct collapse-mitigating structures.

Another interesting feature of these aggregate structures is that there is no adhesion between respective elements. This allows them to respond in drastically different ways to slightly different inputs, and quickly and easily change shape. One could imagine building stable elastogranular structures that could handle large compressive loads, but then be shortened or demolished in seconds with the appropriate input. One example of such a structure is shown in Figure \ref{spacingfig}c. We 3-d printed a plastic (ABS P430) helix which we filled with rocks to form a column. This column was stable in compression, however when the helix was uncoiled, which could easily be done by hand, it collapsed layer-by-layer, allowing for some control over its final height.

\section{Conclusions}
In this work, we set out to understand how elastic rods constrain and jam granular matter. We have found that the elastogranular interaction in this regime is predictable and robust, allowing us to quickly and reliably form structure which can bear significant load (although the limits of this load, and the intricacies of the mechanical response of the columns to compression is beyond our current scope). The stability of these structures depends on the size and frictional properties of the grains, as well as flexibility of the rods. We have so far only considered external loops of elastic, although it has been shown that internal elastic rods are also sufficient to form structure from granular material~\cite{aejmelaeus2017granular}. We expect that this will be related to the mechanics which govern the knotting of ropes~\cite{bayman1977theory, maddocks1987ropes} and, similar to the jamming of chains of beads~\cite{brown2012strain, dumont2018emergent} could be understood through a similar mathematical framework as governs the entanglement of polymers~\cite{doi1996introduction}.

\section*{Conflicts of interest}
There are no conflicts to declare.

\section*{Appendix A: critical spacing}
To find the critical spacing for both experiments and simulation, we build columns with a fixed initial height $H_{i}$ and vary the number of strings with equal spacings $S_{i}$ between them. We linearly fit the data for which $H_f<H_i$ and extrapolate to find the point where $H_f=\gamma H_i$. We use $\gamma = 0.95$ for experiments and $\gamma = 0.9$ for simulations.

The reason that we use a tolerance factor $\gamma$ which is less than 1 is because we found that, in practice, when we removed the slip-cast mold, the grains in the columns had a tendency to settle, and even if $S_i<<S_c$ some of the grains from the top of the columns would dislodge, shortening the column slightly. We chose $\gamma$ empirically as the fraction of $H_i$ that most columns tended to surpass when $S_i<<S_c$.

 \begin{figure}
\begin{center}
\vspace{1mm}
\includegraphics[width=0.98\columnwidth]{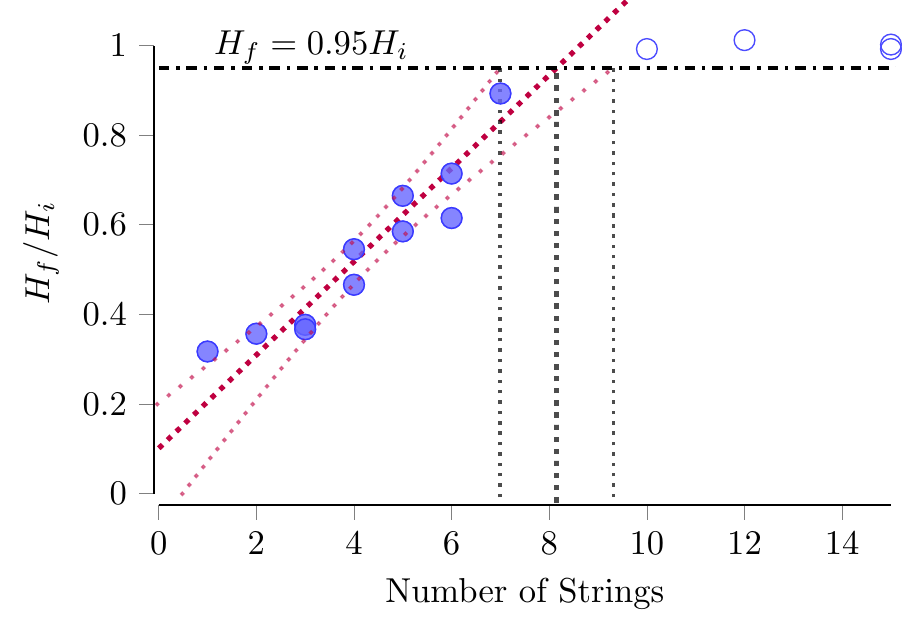}
\end{center}
\vspace{-5mm}
\caption{ To find the critical spacing we plot the final height of the columns vs the number of strings, which is linear when $S_i>S_c$ (filled circles) based on the data presented in Figure~\ref{spacingfig}. We fit this data and find $S_c$ as the spacing at which $H_f>\gamma H_i$ where $\gamma$ is an empirical tolerance factor close to 1. The error on the linear fit gives us the error in $S_c$.
\label{Scfig}}\vspace{-5mm}
\end{figure}

\section*{Appendix B: simulation of loops}

\begin{figure}
\begin{center}
\vspace{1mm}
\includegraphics[width=0.98\columnwidth]{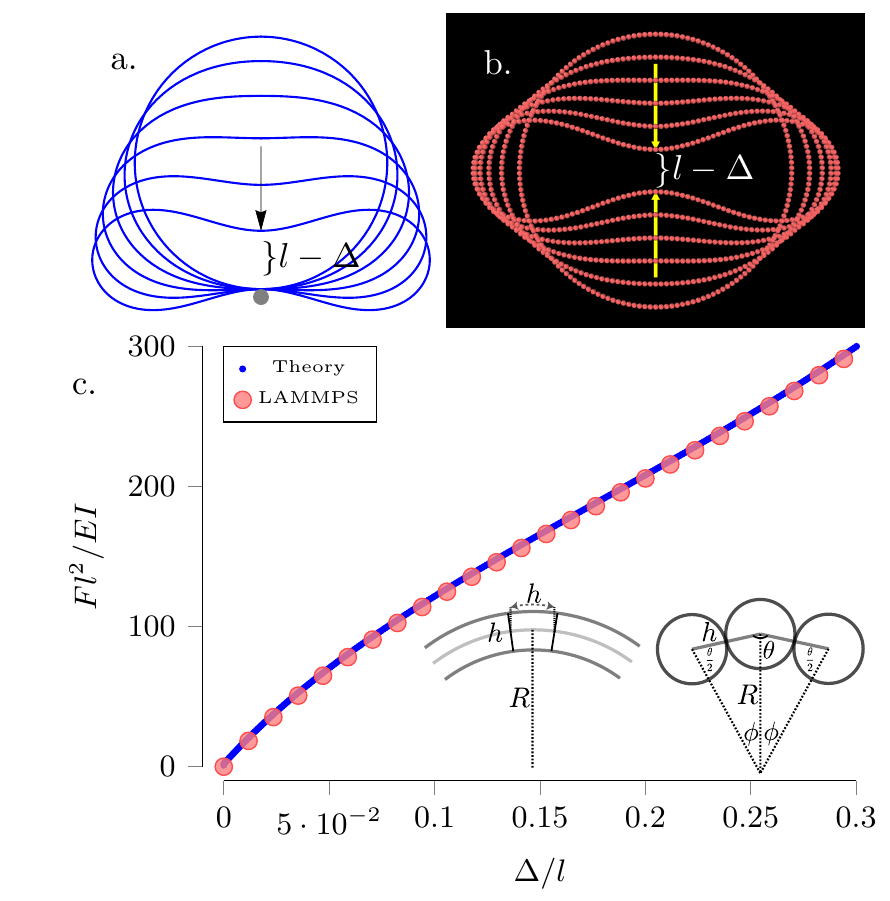}
\end{center}
\vspace{-5mm}
\caption{ (a) Theoretical predictions for the shape of a loop of circumference $l$ loaded with a force $F$ at two points opposite one another. (b) Results of a LAMMPS simulation of the setup in (a). (c) Non-dimensionalized Force $Fl^2/EI$ vs displacement $\Delta/l$ for a theoretical Euler-Bernoulli loop and a loop simulated in LAMMPS. In the inset we illustrate two beams bent with a radius of curvature R -- (left) a continuous cylindrical beam of diameter h, and (right) a discrete cylindrical beam made up of particles of diameter h. The energy of bending of the discrete beam will depend on the angle $\theta$ between each triplet of particles. 
\label{bendingfig}}\vspace{-5mm}
\end{figure}

We would like to replicate the physics of a thin cylindrical loop with Young's modulus $E$, circumference $l$, and cross sectional diameter $h$ (Figure~\ref{bendingfig}a). We will connect particles of diameter $h$ in a loop (Figure~\ref{bendingfig}b) with the following potentials.

To add a stretching rigidity to the loop we will apply a harmonic ``Bond'' between adjacent particles in the beam using the LAMMPS \verb|bond_style harmonic| pair potential. We define an equilibrium distance between adjacent particles, or equilibrium Bond length $r_0$ and calculate the energy of a Bond between particle $i$ and particle $j$ which has a Bond length $r_{ij}$ as 

\begin{equation}
U_s = Y(r_{ij}-r_{0})^2 = Y(\Delta r_{ij})^2
\end{equation}
\label{bondStyle}

Where we define $\Delta r_{ij}$ to be the change from equilibrium of the Bond length. The force on the particles are as such $F=dU_s/dq = 2Y\Delta r_{ij}$. In all of the simulations above, $r_0$ is equal to the diameters of the particles $h$, that is, the potential acts to keep the particles of the beam just in contact. This is not necessary, one could imagine a beam made up of more spherical particles with some neighbor overlap, or fewer spherical particles with a larger distance between each pair.

We take the definition of the Young's modulus $E=\sigma / \epsilon = (F/A)/(\Delta l/l_0)$ where A is the cross sectional area of the loops $\pi (h/2)^2$. If we statically compress or stretch a beam, the force on the ends will be equal to the force between any adjacent particle, which we found above to be $F = 2Y\Delta r$ (we have dropped the subscript of $r$ because the Bond lengths will all be the same). The strain of the whole loop $\Delta l/l_0$ will be equal to the strain of each Bond $\Delta r/r_0$ so we have 

\begin{equation}
E = \frac{F/A}{\Delta l/l_0} = \frac{2Y\Delta r}{\pi (h/2)^2} \frac{h}{\Delta r} = Y \frac{8}{\pi h} \implies Y = \frac{\pi}{8} Eh
\end{equation}

Where we have used the fact that $r_0=h$. 

To derive the pair potential needed to induce the correct bending rigidity in our LAMMPS beams we will consider first a continuous beam (Figure~\ref{bendingfig}c inset, left). We will bend the beam and find the resultant bending energy, and we will use that as the target energy for a discrete beam bent in the same orientation.

For a continuous elastic beam with a moment $I$, the energy due to bending is 

\begin{equation}
U_b^{c} = \frac{EI}{2} \int \kappa^2 ds
\end{equation}

Where $\kappa$ is the curvature of the beam, and we have used the superscript $c$ to indicate that the beam is continuous. To achieve an energy of bending in LAMMPS we will use what is called an ``Angle.'' A LAMMPS Angle is like a Bond (which we used in the previous section on the stretching potential) except instead of considering the interaction between two particles, each Angle applies a potential based on the relative positions of three particles. The specific Angle potential that we will use is \verb|angle_style cosine| which applies a potential

\begin{equation}
U_b^{d} = B (1+\cos\theta_{ijk})
\end{equation}
\label{angleStyle}

Where $\theta_{ijk}$ is the angle between the three particles (Figure~\ref{bendingfig}c inset, right) and we have used the superscript $d$ to indicate that the beam is discrete. We will bend both the continuous and the discrete beams to a radius of curvature $R$ such that for both beams $\kappa = 1/R$ everywhere (Figure~\ref{bendingfig}c, inset). The energy in a chunk of the continuous beam of width $h$ will be 

\begin{equation}
U_b^{c} = \frac{EI}{2R^2} \int ds = \frac{EI}{2R^2} h = \frac{E\pi h^5}{128 R^2}
\end{equation}

Where we have taken the moment of the cylindrical beam $I=\pi h^4/64$. In the discrete case, in a chunk of width $h$ in the bulk of the beam there is one particle (since the particles have width $h$). The energy assigned to that particle because of the bending of the beam can be found by summing up the contributions of each of the Angles that it is a part of. There are three particles in each Angle, so to each constituent particle we will assign one third of the energy in that Angle. Furthermore each particle is a part of three angles, so the energy in each particle in the column due to bending, and therefore the energy in a chunk of the beam of width $h$ is $3*(1/3)B(1+\cos\theta) = B(1+\cos\theta)$ where we have used the fact that the entire beam is bent to the same curvature so each angle $\theta_{ijk}$ between all sets of particles is the same.

The only thing that is left is to connect $(1+\cos\theta)$ to the radius of curvature $R$. We will do this geometrically referring to the angles and lengths defined in Figure~\ref{bendingfig}c inset, right. By the law of cosines, we have that 

\begin{equation}
h^2 = 2R^2-2R^2\cos\phi = 2R^2(1-\cos\phi)
\end{equation}

Since the interior angles of a triangle must sum to 180 degrees, we have that $\phi = 180+\theta/2+\theta/2 \implies \theta = 180-\phi$ so $1+\cos\theta = \frac{h^2}{2R^2}$. Setting the energies equal we have

\begin{equation}
U_{b} = \frac{Bh^2}{2R^2} = \frac{E\pi h^5}{128 R^2} \implies B=\frac{E\pi h^3}{64}
\end{equation}

Thus the total energy of the loops is

\begin{equation}
U = \frac{\pi}{8} Eh \sum (r_{ij}-r_{0})^2 + \frac{E\pi h^3}{64}\sum (1+\cos(\theta_{ijk}))
\label{hopperStable}
\end{equation}

To validate this result we simulate a loop compressed by two point forces and plot the distance between the points $\Delta$ as the force on the loop increases (Figure~\ref{bendingfig}). We find good agreement between the simulation and the theoretical prediction.

\section*{Acknowledgements}
The authors gratefully acknowledge the financial support from DARPA (\#HR00111810004) and from NSF CMMI--CAREER through Mechanics of Materials and Structures (\#1454153), and the computing resources of the Boston University Shared Computing Cluster. We also thank Kate Flanagan and Xin Jiang for initial experimental design and Skylar Tibbits for initial discussions.

\bibliography{Emergence.bib}  

\end{document}